\documentclass[aps,amssymb,physrev,preprint,superscriptaddress,showkeys]{revtex4-2}
\usepackage{graphicx}
\usepackage{amsmath}
\usepackage{dcolumn}
\usepackage{bm}
\usepackage{hyperref}
\usepackage{esvect}

\begin{document}

\title{Influence of the N\'eel vector on the bulk electronic structure of MnTe measured from the \texorpdfstring{$(1 \bar1 00)$}{} surface}

\author{D.A. Usanov}
\affiliation{Laboratory for Quantum Magnetism, Institute of Physics, Ecole Polytechnique Federale de Lausanne, CH-1015 Lausanne, Switzerland}
\email{dmitrii.usanov@epfl.ch}

\author{A.B. Hellenes}
\affiliation{Institute of Physics, Czech Academy of Sciences, Cukrovarnick\'a 10, 162 00 Praha, Czech Republic}

\author{S. Hahn}
\affiliation{MAX IV Laboratory, Lund University, 221 00 Lund, Sweden}

\author{M. Leandersson}
\affiliation{MAX IV Laboratory, Lund University, 221 00 Lund, Sweden}

\author{J. Osiecki}
\affiliation{MAX IV Laboratory, Lund University, 221 00 Lund, Sweden}

\author{K. Uhl\'i\v{r}ov\'a}
\affiliation{Faculty of Mathematics and Physics, Charles University, 121 16 Prague, Czech Republic}

\author{D. Evtushinsky}
\affiliation{Laboratory for Quantum Magnetism, Institute of Physics, Ecole Polytechnique Federale de Lausanne, CH-1015 Lausanne, Switzerland}


\author{L. \v{S}mejkal}
\affiliation{Max Planck Institute for the Physics of Complex Systems, 01187 Dresden, Germany}
\affiliation{Max Planck Institute for Chemical Physics of Solids, 01187 Dresden, Germany}
\affiliation{Institute of Physics, Czech Academy of Sciences,
Cukrovarnick\'a 10, 162 00 Praha 6 Czech Republic}

\author{T. Jungwirth}
\affiliation{Institute of Physics, Czech Academy of Sciences, Cukrovarnick\'a 10, 162 00 Praha, Czech Republic}
\affiliation{School of Physics and Astronomy, University of Nottingham, Nottingham NG7 2RD, United Kingdom}
\affiliation{Center for Science and Innovation in Spintronics, Tohoku University,  980-8577 Sendai, Japan}

\author{J.H. Dil}
\affiliation{Institut de Physique, \'{E}cole Polytechnique F\'{e}d\'{e}rale de Lausanne, CH-1015 Lausanne, Switzerland}
\affiliation{Center for Photon Science, Paul Scherrer Institut, CH-5232 Villigen, Switzerland}

\date{\today}
\begin{abstract}

Altermagnetism gives rise to momentum-dependent spin splitting of the electronic bands without net magnetization. Here, we use angle-resolved photoemission spectroscopy (ARPES) to investigate bulk $\alpha$-MnTe single crystals cleaved along the $(1\bar{1}00)$ plane - an orientation that preserves the compensated magnetic symmetry at the surface and substantially reduces surface effects and $k_z$ broadening. We resolve the three-dimensional bulk band structure with markedly enhanced spectral clarity. The $d$-wave band splitting, driven by the altermagnetic order and spin-orbit coupling in the $\mathbf{\Gamma}-\mathbf{K}-\mathbf{M}$ nodal plane, 
shows a perfect match to theoretical predictions. We further establish that the valence-band maximum occurs near the $\mathbf{A}$ point, identifying the corresponding hole pockets as the states governing the transport properties. Spatially resolved measurements reveal systematic variations of the band dispersion along $\mathbf{\Gamma}-\mathbf{K}$ and $\mathbf{A}-\mathbf{L}$, associated with different pinning of the N\'eel vector at the surface. Our results establish a surface-sensitive route to resolving the bulk altermagnetic electronic structure of MnTe and provide direct spectroscopic evidence for N\'eel-vector-dependent band splitting.

\end{abstract}


\maketitle

\section{\label{sec:Introduction}Introduction} 

Among the materials predicted to host the recently discovered altermagnetic phase \cite{vsmejkal2022beyond, vsmejkal2022emerging}, manganese telluride in the nickel arsenide structure ($\alpha$-MnTe) has attracted considerable attention from both theoretical and experimental perspectives. The main reasons are the relatively simple crystal structure with only two elements, and the well characterized magnetic structure at low temperatures \cite{kunitomi1964neutron, liu2026observation}. The manganese atoms are arranged in two time-reversed sublattices, connected by a non-symmorphic sixfold screw operation along the $c$ axis. This symmetry leads to the lifting of the Kramers spin degeneracy in a form of $g$-wave spin order. In the non-relativistic nodal plane $\mathbf{\Gamma}-\mathbf{K}-\mathbf{M}$ in the Brillouin zone, which corresponds to the crystallographic c-plane, the spin degeneracy is further lifted under the influence of spin-orbit coupling (SOC), giving rise to a characteristic $d$-wave-like spin order \cite{hellenes2024electronic, takahashi2025symmetry, chen2026dominant}.

Angle-resolved photoemission spectroscopy (ARPES), with and without spin resolution, is the most direct probe of the electronic band structure \cite{zhang2022angle, dil2019spin, okuda2017recent}, and has been among the first experimental techniques used to search for signatures of the altermagnetic degeneracy lifting in MnTe. To date, most studies have focused on the $(0001)$ surface, both in thin films \cite{krempasky2024altermagnetic, lee2025dichotomous, lee2024broken, hajlaoui2024temperature, din2025unconventional} and single crystals \cite{krempasky2024altermagnetic, osumi2024observation, martuza2025itinerant, zeng2025non}. However, accessing the bulk electronic structure from this surface with high resolution has proven challenging for several reasons. First, the spectra are affected by surface states and resonances \cite{usanov2026discerning}, likely associated with unsaturated dangling bonds at the surface \cite{he2025epitaxial}. Second, the either Te- or Mn-terminated surface hosts an effectively ferromagnetic Mn layer which breaks the compensated order of the bulk and thereby might influence the surface-sensitive ARPES response. Third, the relatively short $\mathbf{\Gamma}-\mathbf{A}$ distance complicates the observation of band dispersion along this direction in the vacuum ultraviolet (VUV) range \cite{usanov2025spin, din2025unconventional, usanov2026discerning}, because the intrinsic $k_z$ broadening becomes comparable to the Brillouin-zone dimension along this axis. The data interpretation is further complicated by the influence of non-symmorphic symmetry elements, which result in suppression of the photoelectron intensity in every second Brillouin zone along $\mathbf{\Gamma}-\mathbf{A}$ \cite{krempasky2024altermagnetic, usanov2026discerning}.

\begin{figure}[h]
\centering
    \includegraphics[width=0.99\linewidth]{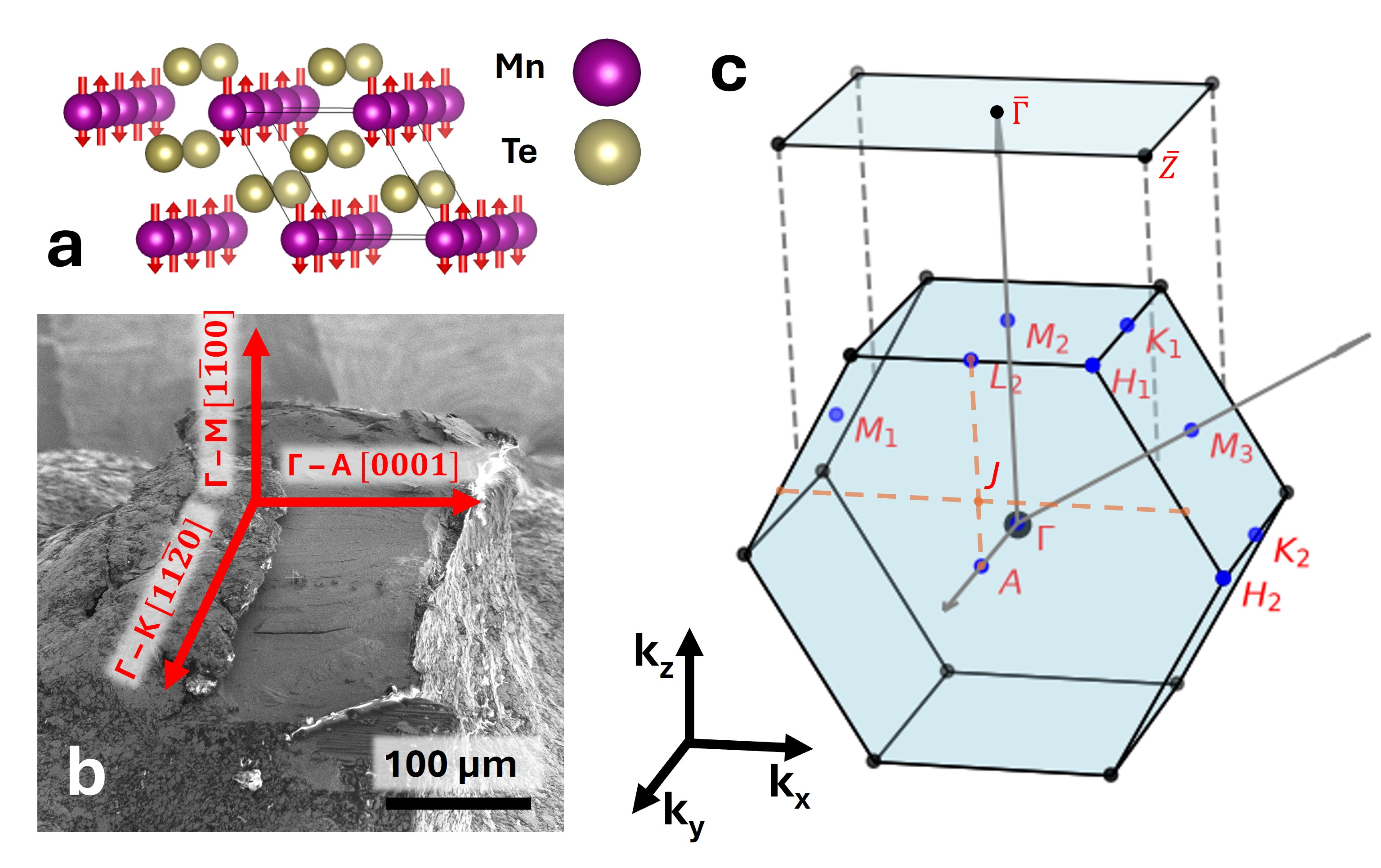}
    \caption{\label{fig:intro} (a) - crystal structure of $\alpha$-MnTe with N\'eel vector along the surface normal and single-layer Te termination. (b) - SEM image of a $(1 \bar1 00)$ cleaved crystal. High symmetry directions in the direct and reciprocal space are indicated. (c) - the bulk and surface Brillouin zones.}
\end{figure}

To overcome these limitations, we performed ARPES measurements on a different surface of MnTe. Experimentally, we identified the $(1 \bar1 00)$ plane as one of the three easy-cleavage planes of MnTe single crystals (Fig.~\ref{fig:intro}). The surface normal $[1 \bar1 00]$ corresponds to the out-of-plane momentum direction $k_z$ along the $\mathbf{\Gamma}-\mathbf{M}$ direction in reciprocal space (Fig.~\ref{fig:intro}b). In contrast to the $(0001)$ surface, this orientation preserves the compensated magnetic order of the bulk, despite the possibility of three distinct surface terminations (Mn-terminated or single-/double-Te-terminated), whereby single Te-terminated is expected to be energetically favorable. Furthermore, the influence of $k_z$ broadening is substantially reduced, because the $\mathbf{\Gamma}-\mathbf{M}$ distance is nearly twice as large as the $\mathbf{\Gamma}-\mathbf{A}$ distance, and for the $\mathbf{\Gamma}-\mathbf{A}$ direction the resolution is only limited by the angular resolution of the set-up. This allows us to characterize the electronic structure with unprecedented quality and show the sensitivity of ARPES to the orientation of the N\'eel vector at the surface.

\section{\label{sec:Methods}Methods}

Single crystals of MnTe were grown using a Te-rich flux. High-purity Mn (99.998\%, Puremat Technologies) and Te (99.9999\%, Mateck) were mixed in a molar ratio of 1:2, loaded into high-purity alumina frit-disc crucibles \cite{Canfield2016}, and sealed in evacuated quartz ampoules. Crystal growth was achieved by cooling from 1050~$^\circ$C to 760~$^\circ$C at a rate of 2.5~$^\circ$C\,h$^{-1}$. The remaining flux was separated from the crystals by centrifugation at 760~$^\circ$C. The samples were etched in aqua regia prepared from HCl (35\%) and HNO$_3$ (65\%) mixed in a 3:1 (v/v) ratio to remove MnTe$_2$ surface impurities and reduce the samples' thickness to approximately 100~$\mu$m.

In order to expose the $(1 \bar1 00)$ surface by cleaving, the crystals were prepared using focused ion beam milling (FIB) \cite{hunter2024controlling} (Helios G4), taking crystal facets and X-ray Laue diffraction data as references. The preparation procedure is detailed in SI.  The orientation of the cleaved surfaces has been reconfirmed using single crystal X-ray diffraction (in SI).

ARPES measurements were performed at the Bloch A end-station of the MAX IV synchrotron facility (Lund, Sweden) \cite{polley2024bloch}, using a DA30-L hemispherical electron analyzer (HSA) from Scienta Omicron, installed with the entrance slit vertically. Angular mapping in the horizontal direction was performed using the deflector. The photon energy range used for this study cover the VUV range from 21 to 190 eV, respective values are specified in the figure captions. Samples were cleaved $in~situ$ and measured at the base temperature of T = 18 K. The angular resolution was approximately 0.2$^o$, and the calculated beamline energy resolution was approximately 13 meV at a photon energy of 99 eV, which was used for most measurements in this study. The light spot size on the sample was $\approx$ 15 $\mu m$ x 10 $\mu m$, comparable to the typical magnetic domain size of the crystals \cite{yamamoto2025altermagnetic, watanabe2026magneto}. The data in the main text is obtained with $p$-polarized light, $s$-polarized light leads to suppression of some bands and is shown in SI.

ARPES data were analyzed and processed using the PESTO Python library \cite{pesto} and using the PEAKS Python package \cite{king2025peaks}. The photon energy scans have been normalized and warped to $k_x$ vs $k_z$ coordinates using the inner potential $V_0$ = 22 eV. The data in the black-blue-white colour scale represent raw intensity, and in black/white - with angular-integrated background subtracted. No additional filtering has been applied.

The electronic band structure calculations were performed with DFT for bulk MnTe, belonging to the space group $P6_3/mmc$, No. 194, with lattice parameters from~\cite{kriegner2017magnetic}. We used the plane-wave Vienna ab initio Simulation Package (VASP)~\cite{kresse1996efficient} with the Perdew-Burke-Ernzerhof (PBE) exchange-correlation functional~\cite{perdew1996generalized} and the projector-augmented-wave method~\cite{blochl1994projector,kresse1999from}, with spin-orbit coupling and electron correlations on Mn $d$ valence electrons using the rotationally invariant Dudarev-type Hubbard correction~\cite{dudarev1998electron} with $U-J=3\,\mathrm{eV}$. We used an $8\times8\times5$ k-point grid, a $520\,\mathrm{eV}$ energy cut-off, and converged the calculations within $10^{-6}\,\mathrm{eV}$.

\section{\label{sec:Results}Results}

\begin{figure}[h]
\centering
    \includegraphics[width=0.99\linewidth]{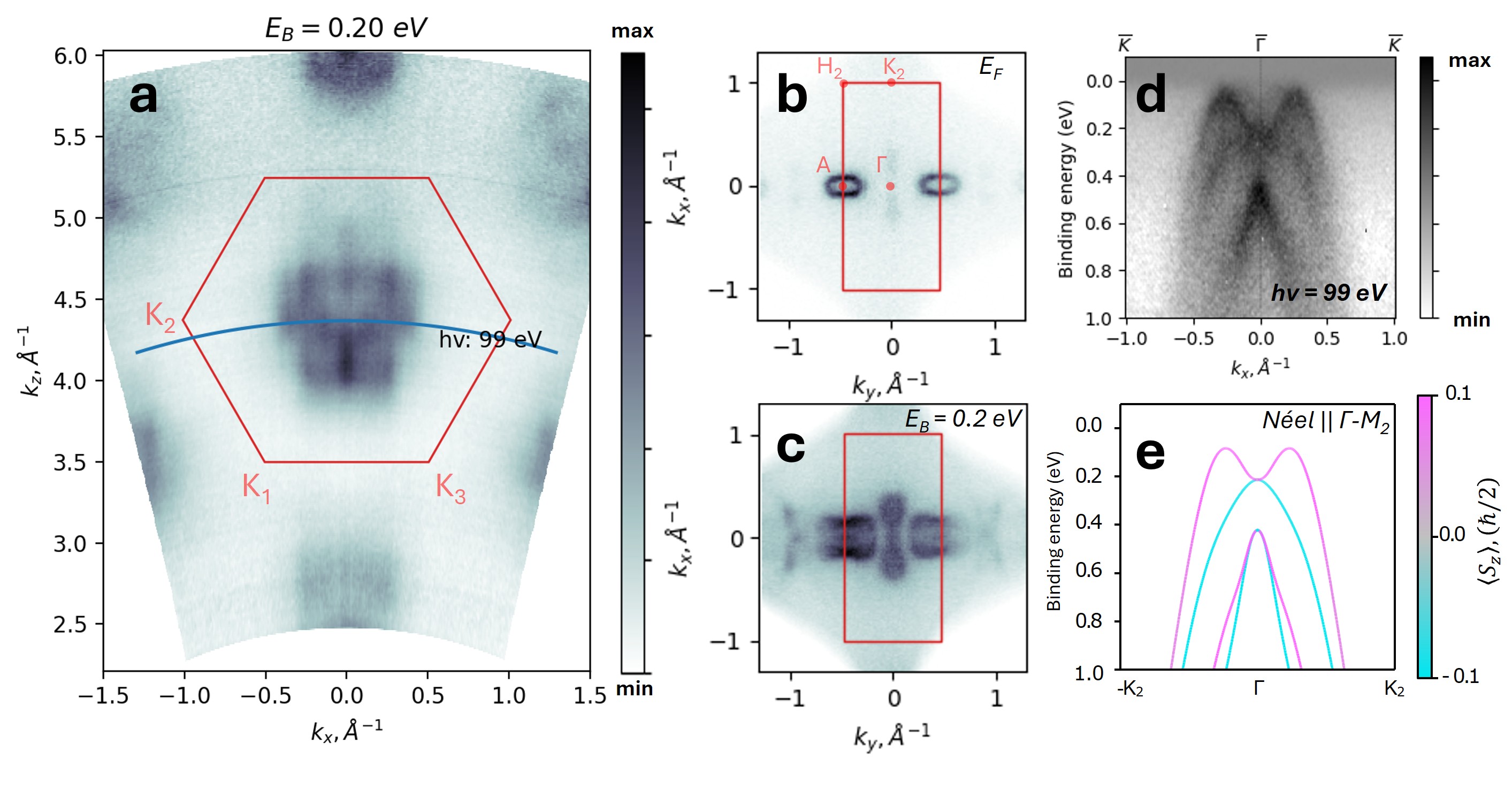}
    \caption{\label{fig:GK}ARPES data obtained with $\mathbf{\overline{\Gamma}}-\mathbf{\overline{K}}$  direction  parallel to the HSA slit. (a) - constant-energy slice at $E_B$ = 200 meV from a photon energy scan at normal emission, position of the ARPES data measured at a fixed photon energy of 99 eV is indicated; (b, c) - constant-energy slices at $E_F$ and $E_B$ = 200 meV from deflector scans at $h\nu$ = 99 eV; (d, e) - Band map along $\mathbf{\overline{K}}-\mathbf{\overline{\Gamma}}-\mathbf{\overline{K}}$  direction at $h\nu$ = 99 eV in comparison with DFT calculations. Borders of the Brillouin zone are highlighted in red.} 
\end{figure}

Fig. \ref{fig:GK} summarizes the main ARPES results obtained with $p$-polarized light and with the $\mathbf{\Gamma}-\mathbf{A}$ direction in the scattering plane (the HSA slit is parallel to the $\mathbf{\Gamma}-\mathbf{K}$). In this fixed experimental geometry, photon energy scans probe the $\mathbf{\Gamma}-\mathbf{K}-\mathbf{M}$ plane of the bulk Brillouin zone, as illustrated in Fig. \ref{fig:GK}a and in SI. Several $\mathbf{\Gamma}$ points can be observed: at $h\nu$ = 50, 99 and 179 eV. The $\mathbf{\Gamma}$ point at around 50 eV is difficult to characterize because it coincides with a photoemission resonance (in SI). It is interesting to note that below this resonance, we do not observe clear angle-dispersing features (in SI). We emphasize that in the full photon energy range, we see no evidences of surface states, in contrast to the $(0001)$ surface. 

At a photon energy of 99~eV, the measured band structure lies approximately in the $\mathbf{\Gamma}-\mathbf{K}-\mathbf{A}$ plane. The corresponding data in Fig.~\ref{fig:GK}b and c, reveal details that were previously inaccessible for the $(0001)$ surface. Along the $\mathbf{\overline{\Gamma}}-\mathbf{\overline{K}}$ direction, a clear band splitting is observed (Fig.~\ref{fig:GK}d), corresponding to the SOC-driven $d$-wave-like altermagnetic splitting at the nodal plane $\mathbf{\Gamma}-\mathbf{K}-\mathbf{M}$, previously referred to as the weak altermagnetic splitting \cite{krempasky2024altermagnetic}. Compared with previous ARPES studies \cite{krempasky2024altermagnetic, osumi2024observation, lee2024broken, hajlaoui2024temperature}, the substantially enhanced spectral resolution of the present measurements enable a considerably sharper and cleaner delineation of the electronic structure.

\begin{figure}[h]
\centering
    \includegraphics[width=0.95\linewidth]{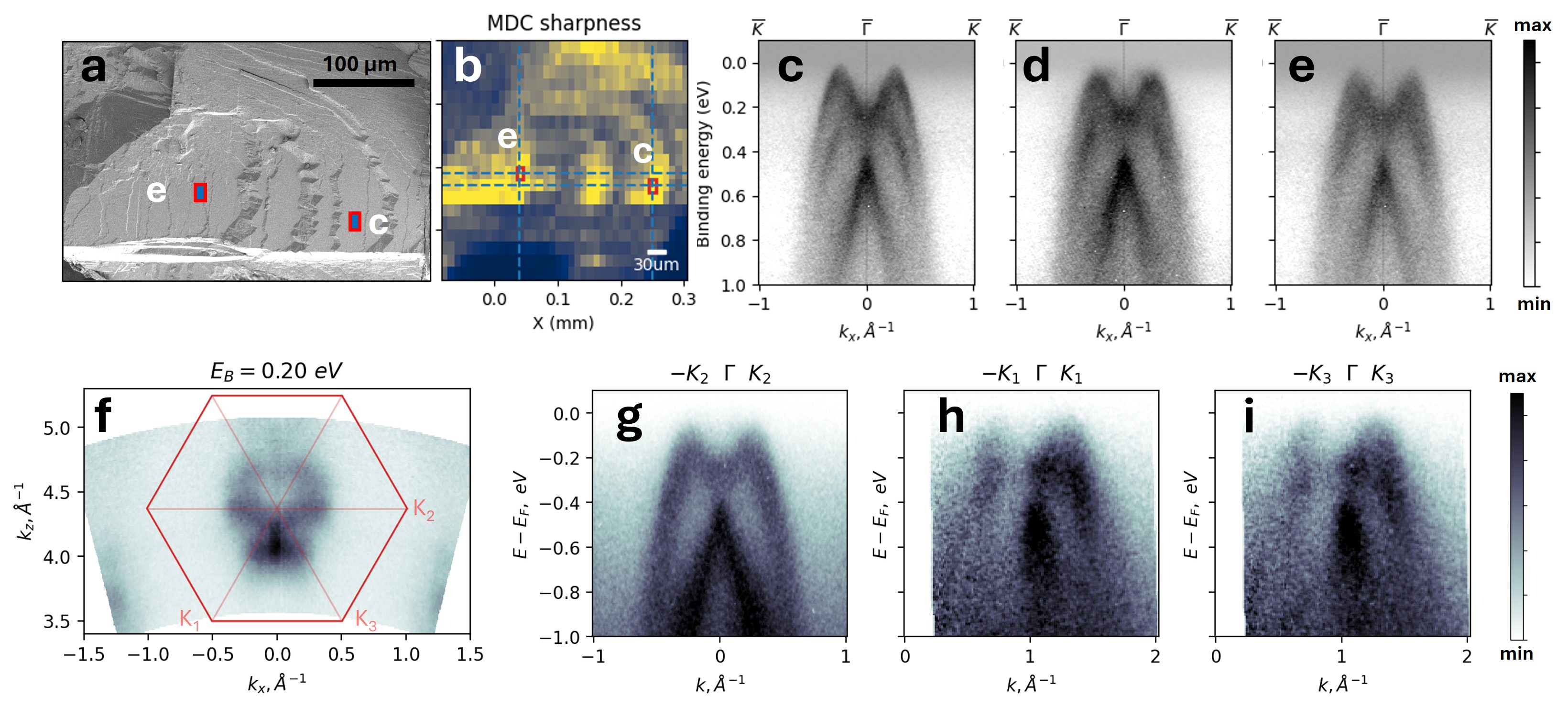}
    \caption{\label{fig:different domains Gamma} (a) - SEM image of the miscut sample; (b) - spatial ARPES image indicating band sharpness at $E_B$ = 200 meV; (c-e) - Band maps parallel to $\mathbf{\overline{K}}-\mathbf{\overline{\Gamma}}-\mathbf{\overline{K}}$ ; (d) - constant-energy slice at $E_B$ = 200 meV from a photon energy scan, taken at the spot (e); (e-g) - corresponding band maps extracted from the photon energy scan along all three $\mathbf{K}-\mathbf{\Gamma}-\mathbf{K}$ directions. Borders of the Brillouin zone are highlighted in red.}
\end{figure}

The high quality of the observed data allows us to rule out the possibility that this splitting originates from a Rashba-like effect. As can be clearly seen in Fig.~\ref{fig:GK}d and Fig.~\ref{fig:different domains Gamma}c--e, the spin split bands follow a parabolic (even-order) dispersion close to $\mathbf{\Gamma}$, whereas bands exhibiting Rashba-like spin textures are expected to display a linear (odd-order) dispersion in this range \cite{hellenes2024electronic, jungwirth2026symmetry}. This confirms the unconventional nature of the observed band splitting and the altermagnetic order of MnTe.

\begin{figure}[h]
\centering
    \includegraphics[width=0.99\linewidth]{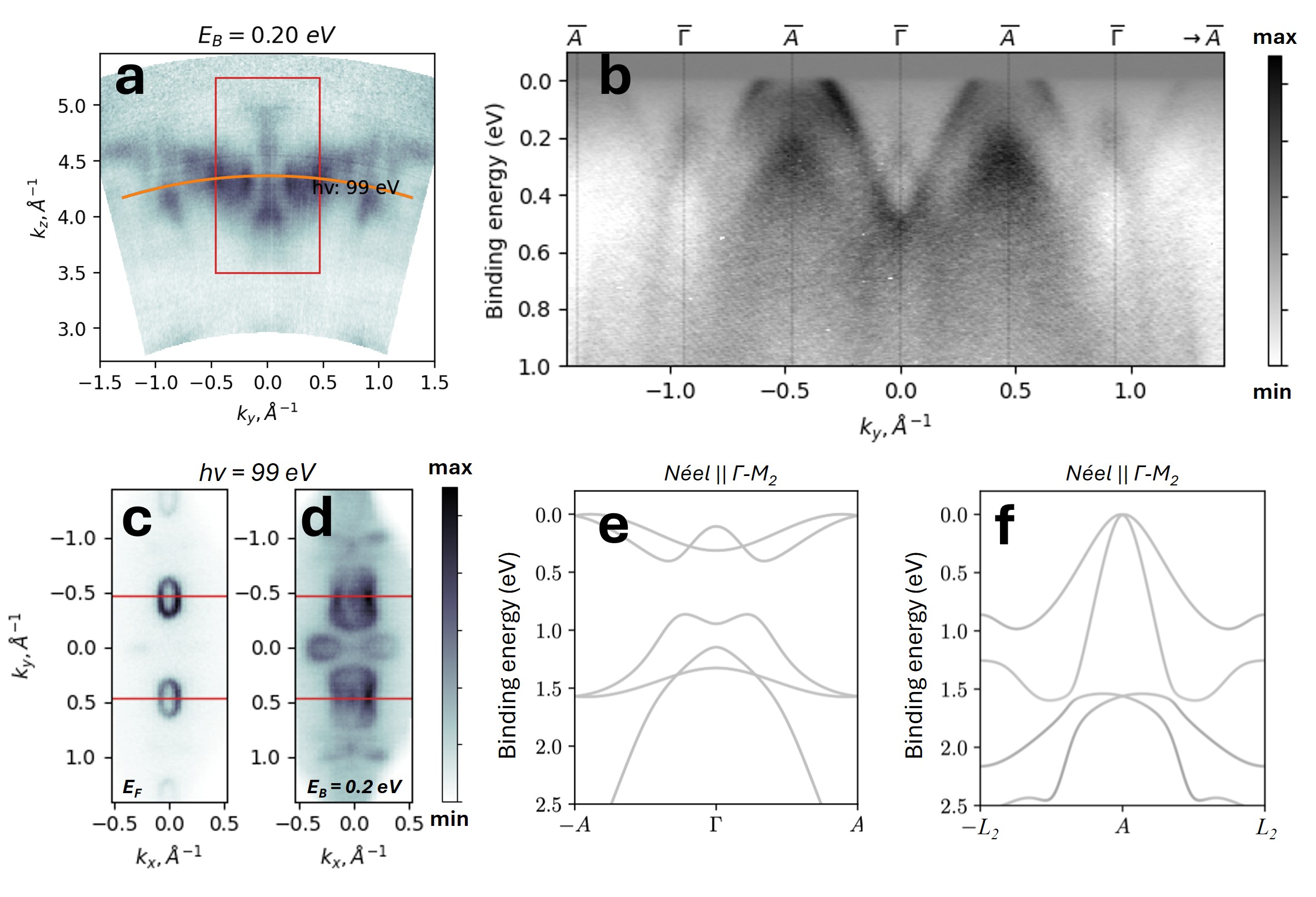}
    \caption{\label{fig:GA} ARPES data obtained with the HSA slit parallel to $\mathbf{\overline{\Gamma}}-\mathbf{\overline{A}}$ direction. (a) - constant-energy slice at $E_B$ = 200 meV from a photon energy scan at normal emission, position of the ARPES data measured at fixed photon energies are indicated; (b) - Energy vs momentum cut along $\mathbf{\overline{A}}-\mathbf{\overline{\Gamma}}-\mathbf{\overline{A}}$  direction at $h\nu$ = 99 eV; (c, d) - constant-energy slices at $E_F$ and $E_B$ = 200 meV from deflector scans at $h\nu$ = 99 eV; (e, f) - calculated band dispersion along $\mathbf{\overline{\Gamma}}-\mathbf{\overline{A}}$ and $\mathbf{\overline{A}}-\mathbf{\overline{L}_2}$. Borders of the Brillouin zone are highlighted in red.} 
\end{figure}

The fact that we can now scan along the $\mathbf{\Gamma}-\mathbf{A}$ direction without changing photon energy, lets us clearly observe hole-like pockets crossing $E_F$, indicating that the top of the valence band is located around the $\mathbf{A}$ point (Fig. \ref{fig:GK}b and Fig.  \ref{fig:GA}b,c). This is even clearer in Fig. \ref{fig:GA}b where the $\mathbf{\Gamma}-\mathbf{A}$ direction is aligned parallel to the HSA slit, i.e., with the sample rotated 90 degrees with respect to the previous alignment. Because of this Fermi level crossing, the transport properties are expected to be governed primarily by these bands rather than by those around $\mathbf{\Gamma}$, meriting closer attention to the region around the $\mathbf{A}$ point, which has so far not been discussed in detail in literature.

Although the observed band dispersion is in a qualitative agreement with the DFT calculations apart from a small energy shift, there are some important differences. The most striking is that in the calculations both spin-degenerate bands are expected to match at the $\mathbf{A}$ point, whereas in the experiment a splitting of around 200 meV is observed. Given the almost perfect match between DFT and experiment around the $\mathbf{\Gamma}$ point, it is possible that this discrepancy is due to a symmetry breaking at the surface, which would otherwise enforce degeneracy at $\mathbf{A}$. 



Because of SOC and the altermagnetic symmetry of MnTe, the magnitude of the band splittings and their spin texture depend on the direction of the N\'eel vector $\vv{N}$. In $(0001)$ surface at low temperature, the easy-axis for $\vv{N}$ is along one of the six $[1 \bar1 00]$ directions, which correspond to $\mathbf{\Gamma}-\mathbf{M}$ directions in reciprocal space. This gives rise to a rich magnetic domain texture, as demonstrated for MnTe thin films \cite{amin2024nanoscale}. Due to strong magnetostiction effects, the degeneracy between the orientational domains can be lifted by micro-patterning of the films, after which a macroscopic single magnetic domain can be created by cooling through $T_N$ in an out-of-plane magnetic field \cite{amin2024nanoscale,din2025unconventional}. In case of the $(1 \bar1 00)$-cleaved crystals, the surface itself breaks the equivalence of the easy-axes in bulk MnTe, which opens a path towards direct observation of the N\'eel-vector-dependent altermagnetic splitting. 

From the experimental point-of-view, discerning N\'eel vector axes requires a beam spot size to be comparable with the size of magnetic domains. Unlike for $(0001)$-oriented pristine thin films, where the characteristic domain size is in 100-nm range, bulk single crystal have magnetic domains in a range from a few $\mu m$ to 100-$\mu m$ scale \cite{yamamoto2025altermagnetic, watanabe2026magneto}. Therefore, the presented measurements performed with a beam spot size around 15$\times$10 $\mu m$, have substantially higher probability to register features of individual domains.

Throughout the measurements, variations of the weak altermagnetic splitting along the $\mathbf{\Gamma}-\mathbf{K}_2$ direction have been observed when probing different spots on the samples. Selected examples are shown in Fig. \ref{fig:different domains Gamma}c-e, and detailed 1D and 2D scans are displayed in the SI. The magnitude of the splitting can be divided into three categories. I: the splitting is the largest and the VBM is close to $E_F$ (Fig. \ref{fig:different domains Gamma}c). II: the splitting is reduced by approx. $50$ meV and the VBM is shifted towards higher binding energy (Fig. \ref{fig:different domains Gamma}d). And III; the spin-split bands come significantly closer to each other, so the splitting is no longer resolved (Fig. \ref{fig:different domains Gamma}e). Exactly these variations have been predicted by bulk DFT calculations (\cite{takahashi2025symmetry} and SI), which consider both easy- and hard-axis directions of the N\'eel vector.

Based on these calculations, and the considerations explained in more detail below, we assign the following N\'eel vectors to the three cases. I: $\vv{N}$ is along the surface normal ($[1\overline{1}00]$ direction). II: $\vv{N}$ is oriented $\pm$60 degrees away from the surface normal (towards the $[21\overline{3}0]$ and $[\overline{1}\overline{2}30]$ directions). III: $\vv{N}$ lies in the surface plane along the $[11\overline{2}0]$ hard axis.

To determine the N\'eel vector direction with better precision, all three $\mathbf{K}-\mathbf{\Gamma}-\mathbf{K}$  directions should be accessed, which is possible to extract from $k$-warped photon energy scans. An example for a selected spot on the sample is shown in Fig. \ref{fig:different domains Gamma}(f-i). It is evident that the band splitting is absent, or minimal, along $\mathbf{\Gamma}-\mathbf{K}_2$, but the $\mathbf{\Gamma}-\mathbf{K}_1$ and $\mathbf{\Gamma}-\mathbf{K}_3$ directions suggest a clear band splitting, which is less visible due to the $k_z$ broadening. This configuration of one small and two large splittings is typical for the N\'eel vector along one of hard axes parallel to one of $[11\overline{2}0]$ ($\mathbf{\Gamma}-\mathbf{K}$) directions (see SI). In this case, the zero splitting is observed for the $\mathbf{\Gamma}-\mathbf{K}_2$ path, indicating the N\'eel vector confinement withing the surface plane. 

Although we see clear variations of the band structure due to magnetic domains, we emphasize that spin-integrated ARPES cannot distinguish domains with opposite directions of the N\'eel vector. It can only determine the axis of the N\'eel vector and not its direction. Spin-resolved ARPES (SARPES) is thus required to fully resolve $\vv{N}$ and, consequently, directly establish the detailed altermagnetic domain structure of MnTe and it's influence on the spin texture. Addressing this question remains an important direction for future studies.

\begin{figure}[h]
\centering
    \includegraphics[width=0.95\linewidth]{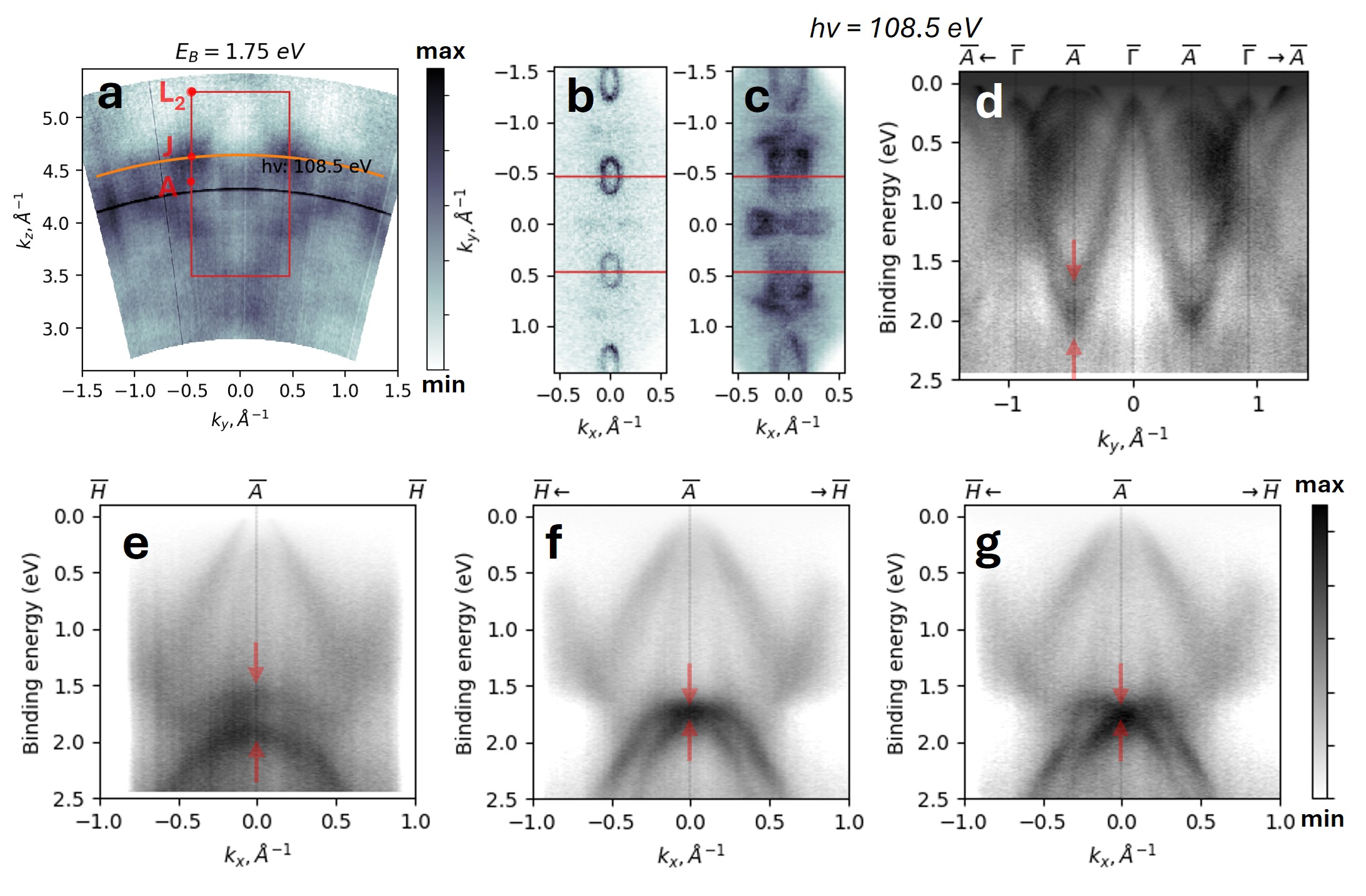}
    \caption{\label{fig: J point} ARPES data obtained with the HSA slit parallel to $\mathbf{\overline{\Gamma}}-\mathbf{\overline{A}}$ direction. (a) - constant-energy slice at $E_B$ = 1.75 eV from a photon energy scan at normal emission, position of the ARPES data measured at fixed photon energy of 108.5 eV is indicated; (b, c) - constant-energy slices at $E_F$ and $E_B$ = 200 meV from deflector scans at $h\nu$ = 108.5 eV; (d, e) - Band map along $\mathbf{\overline{A}}-\mathbf{\overline{\Gamma}}-\mathbf{\overline{A}}$  direction at $h\nu$ = 108.5 eV; (e, f, g) - bandmaps along $\mathbf{\overline{H}}-\mathbf{\overline{A}}-\mathbf{\overline{H}}$ directions at $h\nu$ = 108.5 eV. Borders of the Brillouin zone and the band splittings are highlighted in red.}
\end{figure}

The impact of SOC on the band structure is not limited to the $\mathbf{\Gamma}-\mathbf{K}-\mathbf{M}$ plane. In particular, SOC also lifts the Kramers degeneracy within the $\mathbf{A}-\mathbf{L}-\mathbf{H}$ plane, leaving only $\mathbf{A}-\mathbf{L}$ lines spin-degenerate (in SI), independent of the direction of the N\'eel vector. In ARPES measurements performed from the $(1\bar{1}00)$ surface, the $\mathbf{A}-\mathbf{L}_2$ line can be probed by varying the photon energy at a finite deflection angle from normal emission (Fig. \ref{fig: J point}a). At a particular point along this line, hereafter referred to as $\mathbf{J}$ (Fig.~\ref{fig:intro}b), a pronounced spectral weight is observed at $E_B \approx 1.75$~eV. There the band dispersion varies significantly across different regions of the sample surface. In some regions, two bands separated by approximately 310~meV can be clearly resolved (Fig.~\ref{fig: J point}d,e), whereas in others, the two bands appear to coincide (Fig.~\ref{fig: J point}f,g). We attribute these variations to the different directions of the N\'eel vector. The DFT calculations (in SI) show qualitative agreement with the experimental observations; however, the corresponding variations are less pronounced in the calculations.

In addition to the pinning of the N\'eel vector by the cleaved surface itself, surface roughness may influence the N\'eel vector and hence the observed electronic and magnetic properties. The combined spatially resolved ARPES measurements and scanning electron microscopy in Fig.~\ref{fig:different domains Gamma}a,b, reveal a clear correspondence between surface roughness and the quality of the ARPES spectra. As expected, regions with minimal surface roughness yield the cleanest and sharpest ARPES spectra (region c and Fig. \ref{fig:different domains Gamma}c). In contrast, regions with a higher density of steps exhibit broader bands and additional spectral features (Fig. \ref{fig:different domains Gamma}d and Fig. \ref{fig:GK}d). We thus hypothesize that extended single-terrace regions are more likely to host single magnetic domains. Therefore, experiments aiming to access altermagnetism-specific properties, including spin-resolved ARPES and spectroscopic contrasts such as X-ray circular and linear dichroism, should target flat regions with lateral dimensions comparable to the photon beam spot size. In general, one can assume that the N\'eel vector follows the surface structure on a micron scale \cite{yamamoto2025altermagnetic, watanabe2026magneto, amin2024nanoscale, galindez2025revealing, luo2026symmetry, yamamoto2026nanoscale}, whereas the electronic structure in influenced on a few nm scale.


To conclude, our paper demonstrates a detailed ARPES study of $\alpha$-MnTe, performed on the $in~situ$ cleaved $(1 \bar1 00)$ surface of single crystals. Due to a reduced impact of $k_z$ broadening and a matching compensated magnetic symmetry of the surface and the bulk, surface effects are significantly reduced and therefore the 3D dispersion has been studied with unprecedented accuracy. The non-linear even-parity band dispersion around the $\mathbf{\Gamma}$ point of the spin-split bands confirms the unconventional altermagnetic nature of the splitting and rules out Rashba-like origin. Because of the comparable size of the magnetic domains and the X-ray beam spot, the influence of the N\'eel vector direction on the band structure has been observed, specifically along the $\mathbf{\Gamma}-\mathbf{K}_i$ and $\mathbf{A}-\mathbf{L}_2$ directions. 

\section{\label{sec:Acknowledgments}Acknowledgments}
\begin{acknowledgments}

We acknowledge the MAX IV Laboratory for beamtime on the Bloch beamline under proposal 20252056. Research conducted at MAX IV, a Swedish national user facility, is supported by Vetenskapsradet (Swedish Research Council, VR) under contract 2018-07152, Vinnova (Swedish Governmental Agency for Innovation Systems) under contract 2018-04969 and Formas under contract 2019-02496. We acknowledge the support with SEM and FIB from Lucie Navratilova and the Interdisciplinary Centre for Electron Microscopy at EPFL (CIME). We thank the development community of the PEAKS package \cite{pesto}, PESTO package \cite{king2025peaks} and VESTA \cite{momma2011vesta}. D.A.U. acknowledges support from the Swiss National Science Foundation through the Spark grant CRSK-2 228962. K.U. acknowledges support from the Czech Research Infrastructures program (project No. LM2023065). A.B.H. and T.J. acknowledge support from the European Research Council Advanced Grant No. 101095925. TJ acknowledges support from the MEYS of the Czech Republic Grants No. LM2023051 and CZ.02.01.01/00/22\_008/0004594. We thank Lukasz Plucinski, Eugene Krasovskii, Christophe Berthod for their support with interpreting the results.


\end{acknowledgments}

\bibliography{references}

\end{document}